\documentclass[rnote]{aa}
\usepackage{graphics,epsfig,lscape}
\usepackage{natbib}

\bibpunct{(}{)}{;}{a}{}{,}

\def\cm-2{cm$^{-2}$}

\def\n253{\object{NGC~253}}
\def\m33{\object{M~33}}
\def\mx7{\object{M~33~X$-$7}}
\def\x7{\hbox{X$-$7}}

\begin{document}

\title{Spectroscopy of the bright optical counterparts of X-ray sources in the direction of
          M~31. II}

   \author{P.~Bonfini\inst{1} \and
       D.~Hatzidimitriou\inst{1,3} \and
       W.~Pietsch\inst{2} \and
       P. Reig\inst{3,1}}

       \institute{University of Crete, Department of Physics,P.O.Box 2208, 71003,
       Heraklion, Greece \and Max-Planck-Institut f\"ur extraterrestrische Physik,
           85741 Garching, Germany \and IESL, Foundation for Research and Technology, 71110,
            Heraklion, Greece}

     \offprints{P.~Bonfini, e-mail: {\tt paolo@physics.uoc.gr}}

    \abstract
     {A recent survey of the Local Group spiral galaxy M~31
      with XMM-Newton yielded a large number of X-ray sources.}
     {This is the second in a series of papers with the aim of
     identifying the optical counterparts of these X-ray sources.}
     {We have obtained optical spectra
     for 21 bright optical counterparts of 20 X-ray sources in
     the direction of M~31, using the 1.3-m Skinakas telescope
     in Crete, Greece.}
     {For 17 of the 20 X-ray sources, we have
     identified the optical counterpart as a normal late type
     star (of type F or later) in the foreground (i.e. in the Milky Way).
     For two more sources there were two possible
      optical counterparts in each case, while  two
     more objects have X-ray properties that are not compatible
     with the spectral characteristics of  late type non-flaring
     stars. }{}


   \authorrunning{Optical counterparts of X-ray sources in M~31 field}

\maketitle

\section{Introduction}
We have embarked in an effort of identifying and classifying the
objects responsible for the X-ray emission detected in the
XMM-Newton major-axis survey of M~31 (Pietsch et al. 2005, hereafter
PFH2005). The PFH2005 catalogue includes a total of 856 X-ray
sources, some of which are within M~31 (supernova remnants - SNRs
super-soft sources, X-ray binaries - XRBs and globular cluster
sources, which are most likely low mass X-ray binaries within the
cluster), while others are foreground stars or background active
galactic nuclei (AGNs). A first classification of the sources was
attempted by PFH2005 on the basis of hardness ratio criteria,
however, several sources were not classifiable.

The characterization of the X-ray sources involves optical
spectroscopy of candidate optical counterparts, identified via
cross-correlation with the USNO-B1 catalogue (Monet et al. 2003),
the Local Group Survey Catalogue (LGS, Massey et al. 2006) and the
Two Micron All Sky Survey (2MASS) catalogue (Skrutskie et al. 2006).
The spectroscopic observations are conducted using different sized
telescopes, depending on the magnitude of the candidate optical
counterpart.  In a recent paper (Hatzidimitriou et al. 2006,
hereafter HPM2006) we obtained optical spectra for 14 bright optical
counterparts of X-ray sources in the direction of M~31 (along with
sources in the direction of M33), using the 1.3-m Skinakas
Telescope. The spectra were classified and their spectral classes
compared against their recorded X-ray properties (fluxes and
hardness ratios) to ensure compatibility between the X-ray and
optical characteristics. All 14 objects in the direction of M~31
were confirmed to be foreground stars of spectral types F0 to G9.

In the present paper, we have observed another 21 bright
counterparts of 20 X-ray sources in the direction of M~31 from the
PFH2005 catalogue, down to a limiting magnitude of $R_{lim}=14.3$
using the 1.3-m Skinakas Telescope in Crete (Greece). Of these X-ray
sources, 18 had been classified as foreground star candidates by
PFH2005 on the basis of their X-ray properties and another 2 were
among those classified as `hard' sources (XRBs, Crab-Like SNR or AGN
candidates). It must be noted that none of these bright counterparts
appear in the LGS catalogue, which does not provide photometry for
stars brighter than about $V\sim$15 mag.

In a forthcoming paper (in preparation) fainter optical counterparts
of X-ray sources in the direction of M~31, observed with the 3.5-m
Telescope at Apache-Point Observatory, will be discussed.

In Section 2 we describe the optical observations,  in Section 3 the
data reduction procedure, in Section 4 the spectral classification
method and finally, in Section 5 we discuss the results, including
individual objects of interest.

\section{Optical Observations}
The optical observations used in this study were obtained with the
1.3-m Ritchey-Cretien telescope at Skinakas Observatory, located on
the island of Crete (Greece) and were carried out during three
observing runs, on September 12-15, September 26-30, 2007 and on
October 5-7, 2007.

We used the Low Resolution Spectrograph which is incorporated in the
Focal Reducer instrument. A reflection grating of 1302 lines/mm was
 introduced in the collimator path, giving a nominal dispersion of
 1.04\AA/pixel, and a wavelength coverage from
4680\AA~ to 6761\AA. The spectra were recorded with a 2000x800 ISA
SITe CCD camera. The spectral region was selected  to include both
the H$_{\alpha}$ and H$_{\beta}$ Balmer lines.

Exposure times ranged from 1800~s  to 7200~s, depending on the
magnitude of the object and on seeing and weather conditions. Two or
more exposures were obtained per object. An arc calibration exposure
(NeHeAr) was recorded before and after each object observation.

During the first observing run, 6 targets were observed. During the
second run another 8 targets were observed, of which the candidate
counterpart of source [PFH2005] 766, USNO-B1 1314-0014843, was
resolved into two stars, both within the X-ray source (3-sigma)
error circle. During the final run, 6 more target spectra were
obtained. Thus, spectra for a total of 21 candidate bright
counterparts of 20 PFH2005 sources were acquired. Table \ref{Log}
provides a list of the optical identifications, coordinates and
magnitudes of the objects observed along with a log of the
observations.

\begin{enumerate}

\item {\em Column 1} gives the X-ray source number of the target
as it appears in the corresponding X-ray catalogue paper (PFH2005).
As previously mentioned, for object 766, the bright USNO-B1 star
given as possible optical counterpart in PFH2005  is actually
resolved into  two stars on Skinakas images (with a third just
outside the 3-sigma error circle of the X-ray position). Spectra of
both stars were obtained. Both stars are also listed in the 2MASS
catalogue.

\item {\em Columns 2 \& 3} give the co-ordinates of the X-ray sources
(epoch 2000) as given in the PFH2005 catalogue.

\item {\em Column 4} provides the USNO-B1 identification derived
from the cross-correlation between the X-ray catalogue and the
USNO-B1 catalogue.

\item {\em Columns 5 \& 6} list the $B$ and $R$ magnitudes of the
corresponding USNO object (B2 and R2 in the USNO-B1 catalogue). The
limiting magnitude of the sample of stars observed was $R=$14.3.

\item {\em Column 7} gives the 2MASS identification derived from
the cross-correlation between the X-ray catalogue and the 2MASS
catalogue.

\item {\em Column 8} provides the corresponding $J$ magnitude of the
2MASS counterpart (as given in the 2MASS catalogue).

\item {\em Column 9} provides the corresponding $K$ magnitude of the
2MASS counterpart (as given in the 2MASS catalogue).

\item {\em Column 10} gives the date of the Skinakas observation.

\item {\em Column 11} lists the total exposure time for the
specific spectrum, while the number of exposures taken for that
object is provided in the parenthesis following the total exposure
time.

\item {\em Column 12} gives the spectral type derived in Section 4.

\end{enumerate}

A total of 14 standards of spectral types ranging from O8 to M6 were
observed during the three observing runs in fall 2007, with exactly
the same instrument setup as the targets. The standards were
supplemented by another 15 stars observed with the same instrument
setup in September 2008 with spectral types between F5 and K3. These
complementary data were deemed necessary, as most target stars
proved to be in this spectral range, and a finer spectral-subclass
grid was needed in order to perform accurate spectral classification
(see Section 4). Finally, the 14 standard stars from HPM2006 were
also used, ranging from O9.5 to M6, as they were observed with the
same instrument setup. In this manner, a grid of a total of 43
standard star spectra was constructed.


\begin{table*}
 \begin{center}
  \caption{Log of objects observed in the direction of M~31 and derived spectral types.}
  \label{Log}
  \begin{tabular}{llllllllllll}
   \hline\hline\noalign{\smallskip}
   \multicolumn{12}{c}{{M~31}}\\
   \hline\noalign{\smallskip} {ID}$^1$  &  {RA}   & {Dec} &
   \multicolumn{3}{c} {USNO-B1}  & \multicolumn{3}{c}{2MASS} & {Date} & {Exp.} & {Spectral} \\
   &(2000)&(2000)&{ID}&{$B$}&{$R$}&{ID}&{$J$ }&$K$&&{(s)}&{type}\\
   \noalign{\smallskip} \hline\noalign{\smallskip}

 40      &   00 40 07.29 &   40 41 41.2  & 1306-0011378  &   16.1   &   14.3    &    00400722+4041428   &   13.55  & 13.16 & 29/09/07  &   7200 (4)    &  G2      \\
 42      &   00 40 07.71 &   40 31 12.3  & 1305-0011632  &   13.5   &   12.1    &    00400757+4031134   &   12.18  & 11.85 & 12/09/07  &   3600 (2)    &  F8      \\
 78      &   00 40 40.16 &   40 25 50.9  & 1304-0011697  &   14.4   &   14.2    &    00404007+4025532   &   13.15  & 12.79 & 14/09/07  &   7200 (4)    &  G0      \\
 100     &   00 40 54.54 &   40 31 32.7  & 1305-0012214  &   14.4   &   14.1    &    00405458+4031303   &   13.12  & 12.75 & 06/10/07  &   7200 (4)    &  F7      \\
 128     &   00 41 18.76 &   40 51 59.3  & 1308-001239   &   15.7   &   13.6    &    00411865+4051592   &   11.74  & 11.26 & 29/09/07  &   5400 (3)    &  G6      \\
 129     &   00 41 19.27 &   40 51 01.7  & 1308-0012400  &   13.8   &   13.9    &    00411911+4051017   &   12.01  & 11.48 & 05/10/07  &   5400 (3)    &  K0      \\
 273     &   00 42 29.43 &   41 29 03.0  & 1314-0013202  &   15.2   &   12.4    &    00422944+4129035   &   10.00  &  9.24 & 14/09/07  &   3600 (2)    &  K0      \\
 473     &   00 43 38.58 &   41 37 37.7  & 1316-0014855  &   nf     &   14.2    &    00433861+4137386   &   11.76  & 11.12 & 07/10/07  &   7200 (4)    &  K3      \\
 484     &   00 43 43.10 &   41 38 06.6  & 1316-0014885  &   nf     &   14.2    &    00434293+4138078   &   12.01  & 11.70 & 07/10/07  &   7200 (4)    &  F5      \\
 492     &   00 43 46.69 &   41 38 35.6  & 1316-0014901  &   16.1   &   14.1    &    00434663+4138360   &   11.55  & 11.00 & 07/10/07  &   7200 (4)    &  K2      \\
 593     &   00 44 53.31 &   42 02 15.8  & 1320-0014419  &   14.1   &   13.3    &    00445336+4202145   &   11.76  & 11.16 & 15/09/07  &   2800 (2)    &  K1      \\
 611     &   00 45 05.85 &   42 03 17.3  & 1320-0014588  &   13.5   &   13.1    &    00450551+4203177   &   11.80  & 11.45 & 27/09/07  &   5400 (3)    &  G2      \\
 662     &   00 45 38.97 &   41 56 17.3  & 1319-0014370  &   14.3   &   13.1    &    00453897+4156163   &   11.70  & 11.16 & 30/09/07  &   3600 (2)    &  K0      \\
 663     &   00 45 40.40 &   42 08 07.6  & 1321-0015992  &   13.9   &   11.5    &    00454053+4208066   &   10.46  &    -- & 12/09/07  &   2400 (2)    &  K4      \\
 665     &   00 45 41.26 &   42 19 40.4  & 1323-0017116  &   13.2   &   12.4    &    00454112+4219440   &   11.48  & 11.14 & 15/09/07  &   3600 (2)    &  G0      \\
 701     &   00 46 03.09 &   42 24 34.6  & 1324-0016254  &   14.8   &   14.0    &    00460308+4224338   &   13.08  & 12.68 & 06/10/07  &   7200 (4)    &  F6     \\
 741     &   00 46 23.72 &   41 21 16.7  & 1313-0014047  &   14.6   &   13.7    &    00462372+4121172   &   12.58  & 12.21 & 30/09/07  &   5400 (3)    &  G0      \\
 766$^2$ &   00 46 37.75 &   41 29 05.3  & 1314-0014843  &   14.1   &   12.6    &    00463749+4129053   &   13.56  & 13.23 & 28/09/07  &   3600 (2)    &  G2      \\
         &               &               &               &          &           &    00463781+4129037   &   13.23  & 12.64 & 28/09/07  &   3600 (2)    &  K1      \\
 780     &   00 46 45.85 &   42 30 48.3  & 1325-0017703  &   14.2   &   12.8    &    00464627+4230478   &   11.74  & 11.33 & 27/09/07  &   3600 (2)    &  G2      \\
 840     &   00 47 26.05 &   42 21 58.9  & 1323-0018108  &   13.0    &   12.5    &    00472611+4221583   &   11.27  & 10.92 & 27/09/07  &   3600 (2)    &  G2      \\


   \noalign{\smallskip} \hline\noalign{\smallskip}
   \noalign{\smallskip}
  \end{tabular}
 \end{center}
$^1$ X-ray source numbers from PFH2005 (M~31), i.e. the prefix  $[PFH2005]$
should be added in front of each number.\\

$^2$ The bright USNO-B1 star given as possible optical counterpart
in PFH2005 (USNO-B1 1314-0014843) is actually resolved into two
brighter stars on Skinakas images (and in 2MASS).

\end{table*}



\section{Data reduction}
Data reduction was performed using standard procedures in the
\emph{IRAF} package 2.13-BETA2 (2006). The frames were bias
subtracted, the sky was removed and the  spectra traced and
extracted using the all-in-one subpackage \emph{apextract}. Arc
spectra were extracted from the arc exposures, using exactly the
same profiles as for the corresponding target spectra. The arc
spectra were subsequently used to calibrate the target spectra.
Whenever more than one spectrum of the same object was obtained, the
calibrated spectra were combined, after weighting them according to
exposure time, to yield the final spectrum.


The same reduction and calibration procedure was also applied to the
set of the 43 spectroscopic standard stars. {\footnote  {The spectra
of the standard stars from HPM2006 were re-reduced following the
procedure described here, for consistency, although no differences
in the final spectra were noted.}}

The signal-to-noise ratio (S/N) of the raw target spectra ranged
between $\sim$60 and  $\sim$200. The S/N of the standard star
spectra ranged from $\sim$90 to $\sim$750.

{\bf Figure 1 shows four examples of target spectra. The spectra
were normalized using ordinary \emph{IRAF} procedures.  For
presentation purposes, the wavelength region between 4800-6500~\AA~
is shown. On the same figure we show in red the corresponding
standard star spectra of the same spectral type as the target
spectra, shifted in $y$ by an arbitrary amount.}


\begin{figure*}
\label{overlap}
\resizebox{\hsize}{!}{\rotatebox{0}{\includegraphics{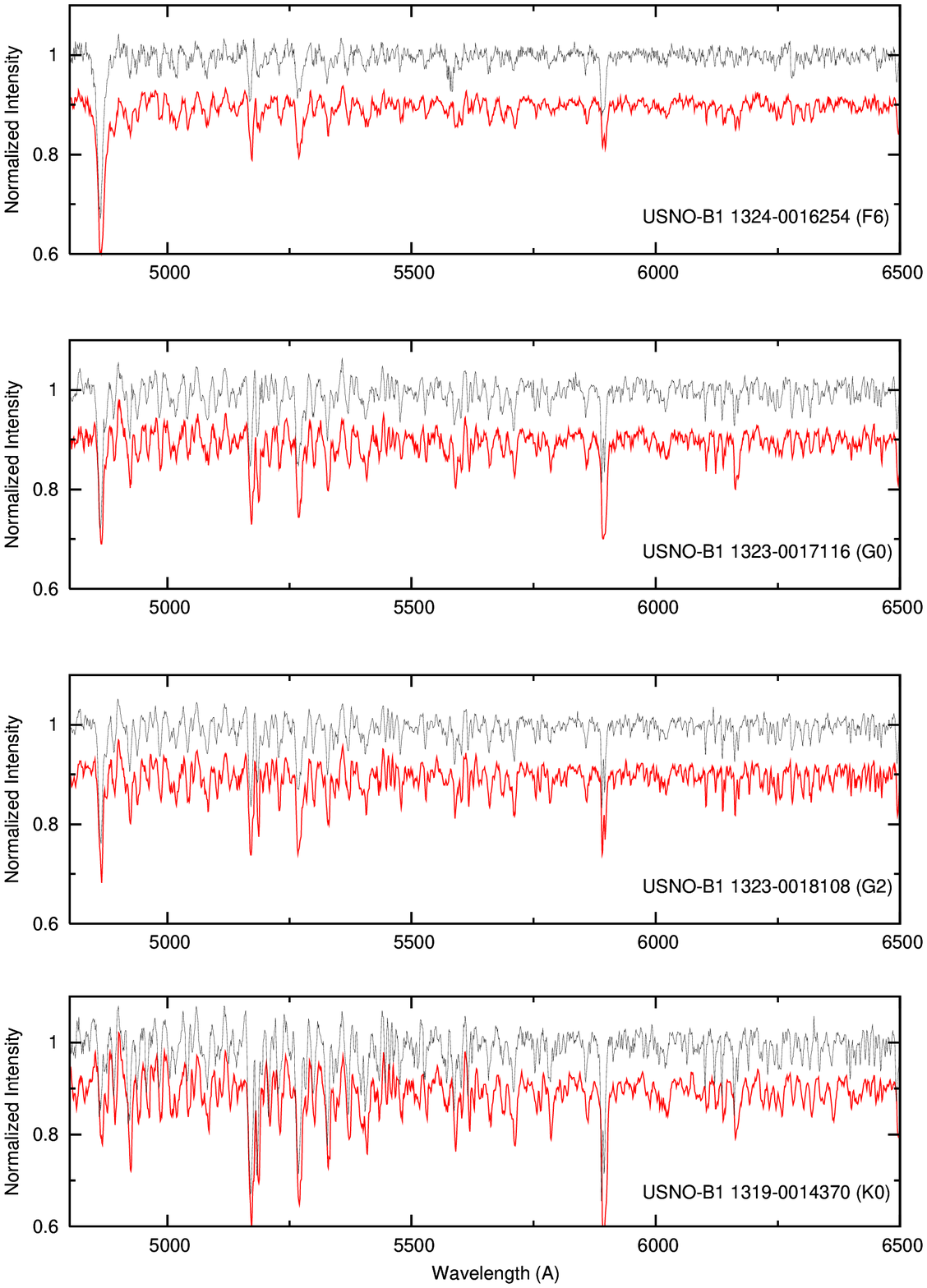}}}
\caption[]{ Four examples of spectra (flux normalized) of optical
counterparts of X-ray sources in the direction of M~31 (black
lines). For comparison, we show in red the corresponding standard
star spectrum (of the same spectral type) shifted in $y$ by an
arbitrary amount. The  wavelength range shown is 4800-6500 \AA.}
\end{figure*}

\section{Spectral classification}

\begin{figure}
\label{PFH05_665}
\resizebox{\hsize}{!}{\rotatebox{0}{\includegraphics{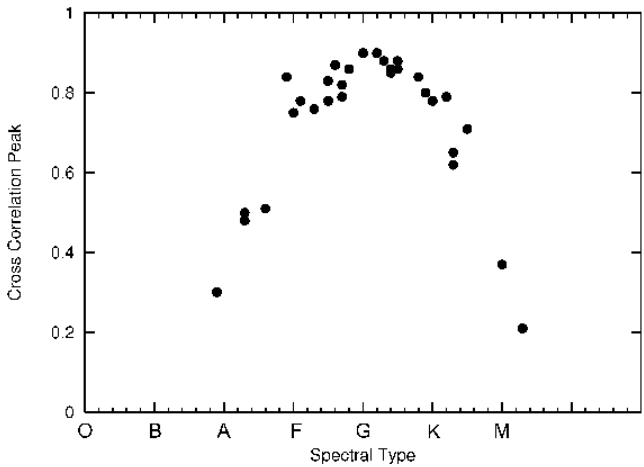}}}
\caption[]{ Results of cross-correlation between the spectrum of
object USNO-B1 1323-0017116 with each standard star spectrum. On the
x-axis, the spectral type of the standard star is indicated, while
on the y-axis the height of the corresponding cross-correlation peak
is given.}
\end{figure}

Classification of the  optical spectra  was achieved via
cross-correlation against the grid of 43 standard star
spectra\footnote{The spectral types we adopted for the standards
were taken from the Catalogue of Stellar Spectral Classification
(Skiff, 2009) table B/mk/mktypes
(http://vizier.u-strasbg.fr/viz-bin/VizieR?-source=B/mk); when a
star had more than one spectral class assigned, we adopted the most
recent one.}, following the procedure described in detail in
HPM2006, i.e. each object spectrum was cross-correlated with each
standard spectrum; the height of the corresponding cross-correlation
peak (hereafter, {\bf CCP}), which is an indicator of the quality of
the match between the target and the standard star spectral
features, was recorded; these {\bf CCP} heights were plotted as a
function of the spectral type of the standard, with the maximum of
the curve yielding the adopted spectral type for the object
spectrum. As an example, Figure \ref{PFH05_665} shows the {\bf CCP}
vs. standard spectral type plot for object USNO-B1 1323-0017116.

As an independent check of the classification process, the spectra
were also visually inspected and compared against the standard star
spectra.

 The  spectral classes derived from the {\bf CCP} vs standard spectral
type plot and from the visual inspection were in very good agreement
in all cases. The final spectral type adopted for each object was
the average of the two estimates ({\bf CCP} and visual) and it is
reported in the last column of Table \ref{Log}.

\begin{figure}
\label{fit_standards}

\resizebox{\hsize}{!}{\rotatebox{0}{\includegraphics{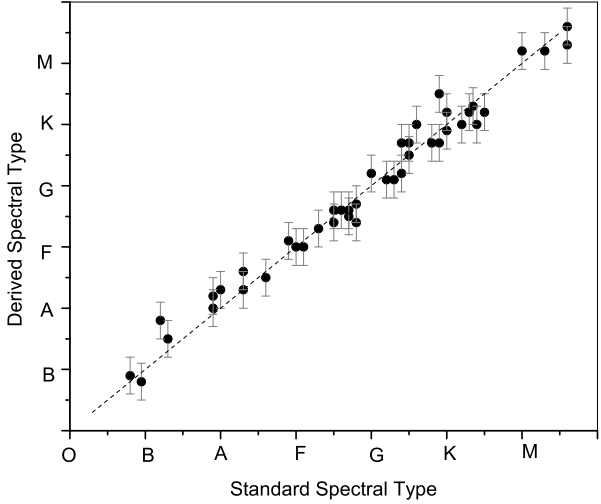}}}
\caption[]{Derived versus standard spectral type for the standard
stars.}

\end{figure}

Conservatively, the accuracy of the adopted spectral classes is
estimated to be $\sim$0.3 of a subclass, following the same
arguments as in HPM2006. It is obvious that the accuracy of the
spectral classification will depend on the fineness of the grid of
standard spectra used, on the signal-to-noise ratio of the
cross-correlated spectra, as well as on spectral class (later
spectral types are easier to classify than earlier types, as
molecular absorption bands are very sensitive to temperature {\bf
and thus the spectrum changes noticeably} between subtypes. However,
these effects were taken into account by making the standard star
grid denser wherever necessary). As in HPM2006, we
 tested the reliability of the adopted classification procedure
by treating each standard in turn as an object spectrum and by
cross-correlating it against the rest of the standards. In Figure
\ref{fit_standards}, we show the derived spectral type for the
standards as a function of the reference spectral type. A
least-squares linear fit yields a slope of $0.95\pm0.03$
(correlation coefficient $R=0.99$), while the average disagreement
between the standard and derived spectral type is 0.23 of a spectral
type.

In Figure 4, we plot the derived spectral types of the optical
counterparts observed against their $J-K$ colour as given in  the
2MASS catalogue. The correlation is quite tight {\bf (correlation
coefficient $R=0.94$, excluding USNO-B1 1314-0013202, see below)},
and in all but one case the $J-K$ colour is in excellent agreement
with the derived spectral type. The one exception is the spectrum of
the counterpart of object [PFH2005] 273 (USNO-B1 1314-0013202) which
has a spectral type $\sim$K0, while its $J-K$ colour suggests type
M. It must be noted that the star is very bright, and  its recorded
magnitude and colour in 2MASS may have been affected by saturation
or linearity effects.

\begin{figure}
\label{JKvsSPTYPE}
\resizebox{\hsize}{!}{\rotatebox{0}{\includegraphics{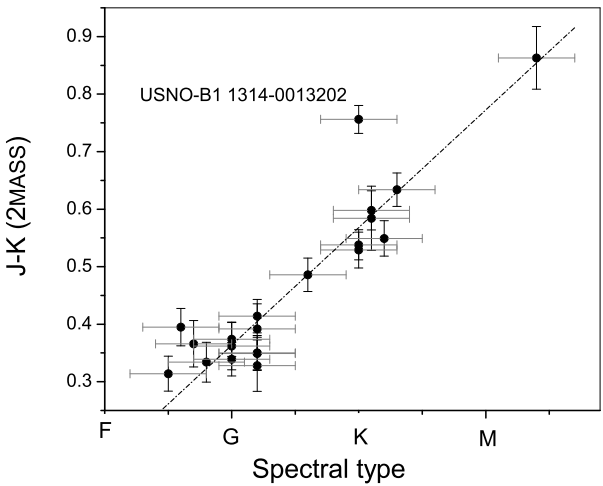}}}
\caption[]{ Derived spectral type versus J-K colour from 2MASS
whenever available.  }
\end{figure}

\section{Discussion and Conclusions}

\begin{figure}
\resizebox{\hsize}{!}{\rotatebox{0}{\includegraphics{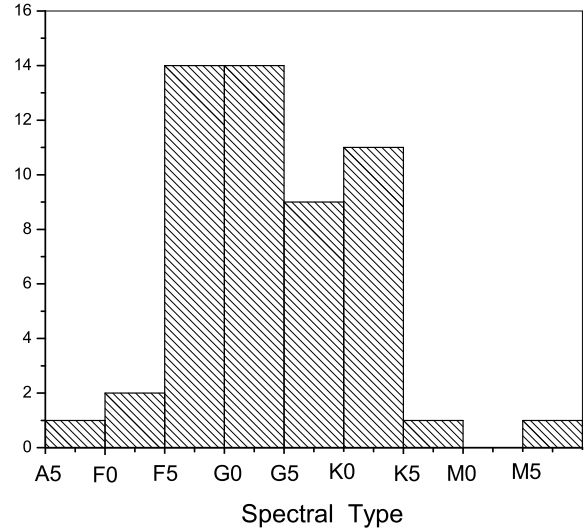}}}
\caption[]{ Frequency distribution of spectral types of the observed
bright stellar counterparts of X-ray sources in the direction of
M~31 and M~33. Results from the present work are integrated with
data from HPM2006} \label{distribution}
\end{figure}

All spectra obtained correspond to foreground stars with spectral
types F5 to K4. None of the stars showed emission in H-alpha, that
could be associated to flaring activity. Figure \ref{distribution}
shows the frequency distribution of the derived spectral types, both
from the present work and from HPM2006. Spectral types peak between
G0 and G5, with all but one object of type F or later. Such later
type stars are known to be soft X-ray emitters due to coronal
activity (e.g. Schmitt 1999, {\bf Schachter et al. 1996, Sciortino
et al. 1993}).

In the case of [PFH2005] 766, two late type foreground stars of
similar optical magnitude lie within the error circle of the X-ray
source (see Table 1). Since any of these stars could be the source
of the X-ray emission, we could not identify with certainty the
optical counterpart. It is also possible that the X-ray emission
results from more than one star, in this case.

In the case of [PFH2005] 663, the optical counterpart USNO-B1
1321-0015992 (which is also recorded as a single star in 2MASS) is
actually resolved into two close bright stars in the LGS catalogue.
The two stars are also resolved on the Skinakas acquisition images,
but the spectra could not be separated. The obtained -presumably
composite- spectrum does not show any indication of the presence of
two separate spectral components and was consistent with a K-type
star.

In order to ascertain that the optical counterparts observed are
indeed responsible for the X-ray emission, compatibility with the
X-ray properties should be examined. In Table \ref{X-ray}, we list
the main X-ray properties of the sources (as given in PFH2005),
namely the X-ray flux, $f_x$, the logarithm of the ratio of the
X-ray flux to the optical flux, $log(f_x/f_{opt})$ (calculated as
described in HPM2006) and  the hardness ratios HR1, HR2, HR3 and HR4
(for the definition of the hardness ratios see PFH2005).

{\bf The logarithm of the ratio of X-ray flux to optical flux,
$log(f_x/f_{opt})$, ranges between $-3.5$ and $-2.8$ for the F stars
in the sample, between $-4.1$ and $-2$ for the G stars and between
$-3.8$ and $-2.4$ for the K stars. These are consistent with the
values expected for normal stars of types F-M (Schachter et al.
1996, Maccacaro et al. 1988, Vaiana et al. 1981). One of the G
stars, the G6 star USNO-B1 1308-001239, has a relatively high value
of $log(f_x/f_{opt})=-2.$, which places it at the edge of the
expected range for G stars (Schachter et al. 1996). There is no
clear correlation between  $log(f_x/f_{opt})$ and spectral type,
while the  ranges of the possible values of $log(f_x/f_{opt})$ for
the different spectral types largely overlap. Indeed, the soft X-ray
emission caused by stellar coronal activity varies significantly
within the same spectral type. This is a well known fact, which
indicates that X ray coronal activity is the result of a combination
of several factors that determine the efficiency of stellar coronal
heating, such as stellar magnetic field, stellar rotation rate,
stellar age and depth of the convection zone (see e.g. Schachter et
al. 1996).}

\begin{figure}
\resizebox{\hsize}{!}{\rotatebox{0}{\includegraphics{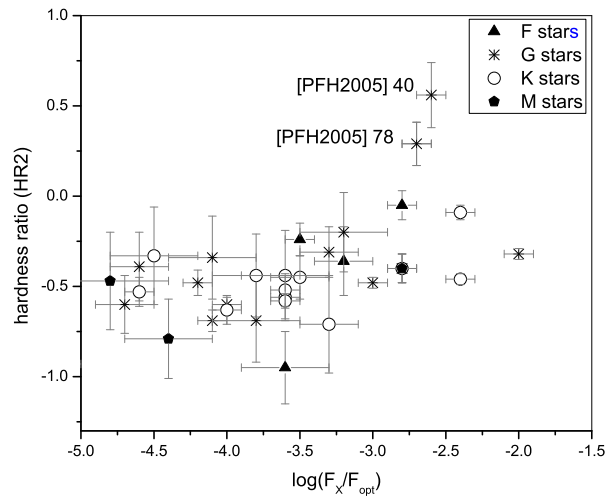}}}
\label{HR2} \caption[]{ X-ray properties of the sample considered in
the present work and in HPM2006. Different spectral types are marked
in different colours. }
\end{figure}

Figure \ref{HR2} shows the hardness ratio HR2 against the X-ray to
optical flux ratio [$log(f_x/f_{opt})$], marking the different
spectral classes in different colours. In the same figure we have
included foreground stars in the HPM2006 sample, excluding object
[PFH2005] 464 with poorly defined hardness ratios (see discussion in
HPM2006). All spectral types occupy  the same locus on the
$log(f_x/f_{opt})$-HR2 plane. The objects appear to lie on a
horizontal branch up to $log(f_x/f_{opt})<$-3.6; beyond this value
there seems to be a small increase in hardness ratio. This behavior
is similar for all the spectral types.  The observed dispersion of
the data points is expected to be affected by the unspecified
luminosity class of the objects, as giants have on average higher
hardness ratios and higher X-ray fluxes than dwarfs. Moreover high
rotation can also affect the location of a star in the above
diagram. Two objects, [PFH2005] 40 and [PFH2005] 78, have hardness
ratios (HR2) that appear to be too high for a normal late type star.
Indeed, both sources were classified as "hard" by PFH2005. Careful
inspection of the X-ray data did not provide any possible cause for
an inadvertent increase of the  hardness ratios recorded (such as
close vicinity to a bright hard source), or show any X-ray
variability. Moreover none of the stars show flaring activity in
their optical spectra.


In the case of [PFH2005] 40, we observed the candidate counterpart
USNO-B1 1306-0011378.  There is no other optical source within the
3-sigma error radius of the X-ray source position in any of the
USNO-B1, 2MASS or the LGS catalogues. The closest neighbour is the
LGS object J004007.76+404140.0 at a distance of 5.48$\arcsec$
South-East of the X-ray position and beyond the 3 sigma error
radius,  at a magnitude of $V=21.6$mag and colour $B-V=1.37$. It is
not clear that this object is a likely counterpart of [PFH2005] 40.
There may be other sources within the 3-sigma error circle that are
too faint to be recorded even in the deepest optical catalogue
currently available. {\bf Figure 7 shows the location of the X-ray
source [PFH2005] 40 along with the 3-sigma error circle, overlayed
on a LGS $R$-band image. The locations of USNO-B1 1306-0011378 and
J004007.76+404140.0 are also marked.}

In the case of [PFH2005] 78, we observed the candidate counterpart
USNO-B1 1304-0011697. Within the 3-sigma X-ray error circle, at a
distance of 2.33 $\arcsec$ South-West of the X-ray position, there
is a $V=$ 21.3mag ($B-V=0.84$) object in the LGS catalogue (object
ID: J004040.03+402549.1). Follow-up spectroscopy of this fainter
object is necessary, in order to proceed with the identification of
the counterpart of this sources. {\bf Figure 8 shows the location of
the X-ray source [PFH2005] 78 along with the 3-sigma error circle,
overlayed on a LGS $R$-band image. The locations of USNO-B1
1304-0011697 and J004040.03+402549.1 are also marked.}


\begin{table*}
 \begin{center}
  \caption{{X-ray properties of the objects of Table \ref{Log}}}
  \label{X-ray}
  \begin{tabular}{lcccccc}

   \hline\hline\noalign{\smallskip} \noalign{\smallskip}
   \multicolumn{7}{c}{{M~31}}\\
   \noalign{\smallskip}\hline\noalign{\smallskip}
{ID}$^1$ &{$f_x$ (mW/m$^2$)}&{{\bf$log(f_x/f_{opt})$}}&{HR1} &{HR2} &{HR3} &{HR4}\\
  \noalign{\smallskip} \hline\noalign{\smallskip}

 40  & 8.4e-15$\pm$1e-15 &     -2.6$\pm$0.1 & 0.12$\pm$0.34 &  0.56$\pm$0.18 & -0.24$\pm$0.18 & -0.45$\pm$0.37 \\
 42  & 1.0e-14$\pm$1e-15 &     -3.5$\pm$0.1 & 0.53$\pm$0.09 & -0.24$\pm$0.09 & -0.27$\pm$0.13 & -0.59$\pm$0.25 \\
 78  & 1.6e-14$\pm$2e-15 &     -2.7$\pm$0.1 & 0.70$\pm$0.18 &  0.29$\pm$0.12 & -0.22$\pm$0.12 & -0.35$\pm$0.27 \\
 100 & 4.8e-15$\pm$1e-15 &     -3.2$\pm$0.2 & 0.59$\pm$0.18 & -0.36$\pm$0.19 & -0.43$\pm$0.30 & -          \\
 128 & 6.1e-14$\pm$2e-15 &     -2.0$\pm$0.1 & 0.56$\pm$0.03 & -0.32$\pm$0.03 & -0.84$\pm$0.04 & -0.97$\pm$0.26 \\
 129 & 3.0e-15$\pm$4e-16 &     -3.6$\pm$0.1 & 0.32$\pm$0.12 & -0.56$\pm$0.12 & -0.78$\pm$0.30 &  -         \\
 273 & 5.0e-14$\pm$2e-15 &     -2.4$\pm$0.1 & 0.51$\pm$0.05 & -0.09$\pm$0.04 & -0.58$\pm$0.05 & -0.62$\pm$0.21 \\
 473 & 1.2e-15$\pm$3e-16 &     -3.6$\pm$0.3 & 0.27$\pm$0.23 & -0.44$\pm$0.25 &  -         & -          \\
 484 & 1.5e-15$\pm$4e-16 &     -3.6$\pm$0.3 & 0.04$\pm$0.27 & -0.95$\pm$0.20 &  -         & -          \\
 492 & 2.1e-15$\pm$5e-16 &     -3.3$\pm$0.2 & 0.28$\pm$0.14 & -0.71$\pm$0.27 &  -         & -          \\
 593 & 2.5e-15$\pm$7e-16 &     -3.8$\pm$0.3 & 0.62$\pm$0.22 & -0.44$\pm$0.23 &  -         & -          \\
 611 & 3.1e-15$\pm$8e-16 &     -3.8$\pm$0.3 & 0.54$\pm$0.21 & -0.69$\pm$0.23 &  -         & -          \\
 662 & 4.2e-15$\pm$7e-16 &     -3.5$\pm$0.2 & 0.59$\pm$0.13 & -0.45$\pm$0.12 & -0.50$\pm$0.24 & -          \\
 663 & 1.3e-13$\pm$5e-15 &     -2.4$\pm$0.1 & 0.36$\pm$0.03 & -0.46$\pm$0.03 & -0.83$\pm$0.05 & -0.32$\pm$0.43 \\
 665 & 2.6e-15$\pm$8e-16 &     -4.1$\pm$0.3 & 0.51$\pm$0.23 & -0.34$\pm$0.23 & -0.83$\pm$0.48 & -          \\
 701 & 1.2e-14$\pm$1e-15 &     -2.8$\pm$0.1 & 0.42$\pm$0.09 & -0.05$\pm$0.08 & -0.63$\pm$0.12 & -0.43$\pm$0.44 \\
 741 & 5.2e-15$\pm$1e-15 &     -3.3$\pm$0.2 & 0.67$\pm$0.18 & -0.31$\pm$0.14 & -0.92$\pm$0.24 & -          \\
 766 & 3.2e-14$\pm$4e-15 &     -2.8$\pm$0.1 & 0.53$\pm$0.08 & -0.40$\pm$0.08 & -0.59$\pm$0.17 & -          \\
 780 & 1.1e-14$\pm$3e-15 &     -3.2$\pm$0.3 & 0.72$\pm$0.19 & -0.20$\pm$0.22 & -0.60$\pm$0.31 & -          \\
 840 & 3.5e-14$\pm$1e-15 &     -3.0$\pm$0.1 & 0.70$\pm$0.03 & -0.48$\pm$0.03 & -0.90$\pm$0.05 & -          \\

   \noalign{\smallskip} \hline\noalign{\smallskip}
  \end{tabular}
 \end{center}
 $^1$ X-ray source numbers from PFH2005 (M~31), i.e. the prefix  $[PFH2005]$ should be added in
 front of each number.\\
\smallskip
\end{table*}

\begin{figure}
\resizebox{\hsize}{!}{\rotatebox{0}{\includegraphics{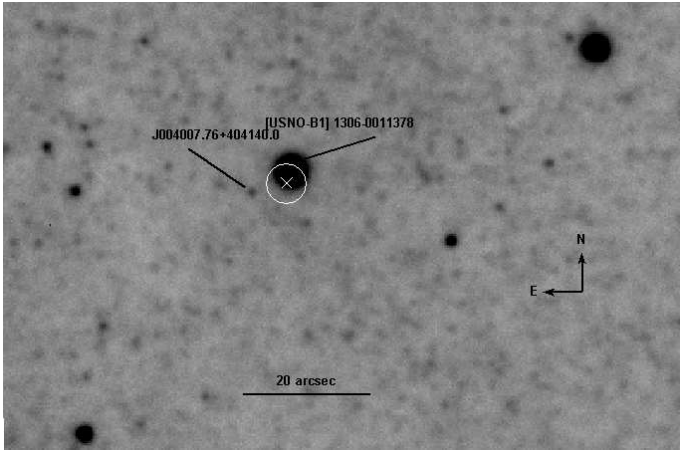}}}
\label{paper_PFH40} \caption[]{ X-ray position shown with a cross
(99.73\% confidence error circle) of object [PFH2005] 40 overlaid on
a R-band optical image (Massey et al. 2006). The bright stellar
object with spectroscopic follow-up is USNO-B1 1306-0011378. The
position of the Massey et al. (2006) catalogue object
J004007.76+404140.0 is also indicated. }
\end{figure}

\begin{figure}
\resizebox{\hsize}{!}{\rotatebox{0}{\includegraphics{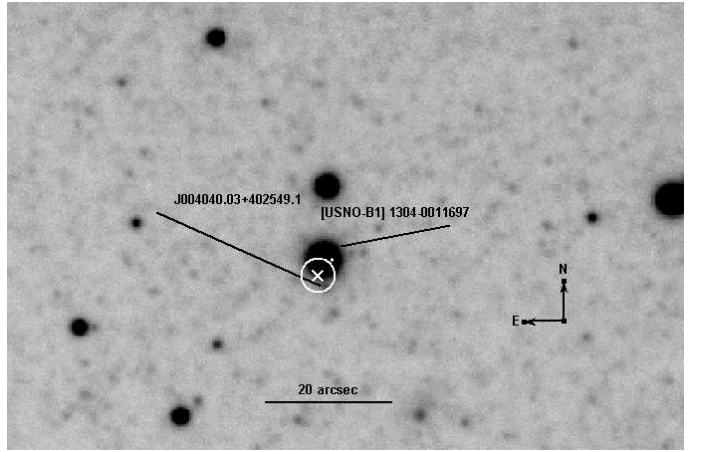}}}
\label{paper_PFH78} \caption[]{ X-ray position (99.73\% confidence
error circle) of object [PFH2005] 78 overlaid on a V-band optical
image (Massey et al. 2006). The bright stellar object with
spectroscopic follow-up is  USNO-B1 1304-0011697. The position of
the Massey et al. (2006) catalogue object J004040.03+402549.1 is
also indicated.
 }
\end{figure}

In conclusion, we have obtained optical spectra for 21 bright
optical counterparts of 20 X-ray sources in the direction of M~31,
using the 1.3-m Skinakas telescope in Crete, Greece. For 17 of the
20 X-ray sources, we have identified the optical counterpart as a
normal late type star (later than type F) in the foreground (i.e. in
the Milky Way). For two more objects, there are two normal late type
stars within the X-ray error circle, and therefore it was not
possible to identify with any certainty the actual counterpart of
the X-ray emission. Two more objects have X-ray properties that are
not compatible with the spectral characteristics of a late type
non-flaring star. In one of these cases, a fainter optical source
within the error circle of the X-ray position is a possible
counterpart, that needs to be confirmed.

\acknowledgements The authors thank T. Koutentakis, N. Primak and V.
 Antoniou who helped with the observations at Skinakas Observatory.

\end{document}